\documentclass[prd,aps,showpacs,tightenlines,nofootinbib]{revtex4}
\usepackage{graphicx}

\newcommand{\bear}{\begin{array}}  \newcommand{\eear}{\end{array}}
\newcommand{\bea}{\begin{eqnarray}}  \newcommand{\eea}{\end{eqnarray}}
\newcommand{\beq}{\begin{equation}}  \newcommand{\eeq}{\end{equation}}
\newcommand{\bef}{\begin{figure}}  \newcommand{\eef}{\end{figure}}
\newcommand{\bec}{\begin{center}}  \newcommand{\eec}{\end{center}}
\newcommand{\non}{\nonumber}

\newcommand{\bib}{\bibitem}

\def\PLB#1#2#3{Phys. Lett. B {\bf #1}, #2 (19#3)}

\def\PRD#1#2#3{Phys. Rev. D {\bf #1}, #2 (19#3)}

\def\PRL#1#2#3{Phys. Rev. Lett. {\bf#1}, #2 (19#3)}

\begin{document}

\title{Inflation model with lower multipoles of the CMB suppressed}
\author{Masahiro Kawasaki  and  Fuminobu Takahashi}
\affiliation{Research Center for the Early Universe, University of
Tokyo, Tokyo 113-0033, Japan}
\date{\today}
\begin{abstract}
The recent observation of the cosmic microwave background anisotropy by
the WMAP confirmed that the lower multipoles are considerably suppressed.
From the standpoint of the cosmic variance, it is nothing but a statistical accident.
Alternatively, one can attribute the deficit of fluctuation on the large scale
to the cosmic history, which might be explained in the context
of the inflationary physics.  In this paper, we show that
it is possible to explain the suppressed lower multipoles in  the hybrid new inflation model. 
\end{abstract}

\pacs{98.80.Cq} \maketitle

The recent observational results of the cosmic microwave background (CMB) anisotropy by the WMAP strongly support the inflationary scenario, on the ground
that the WMAP data are well described by pure adiabatic fluctuations~\cite{
WMAP}.  Although the result agrees with the generic predictions of inflationary scenario within a statistical error, it still suggests two unusual features.
One is the running spectral index, and the other is anomalously low value of the
quadrupole moment of the CMB. In this paper we would like to focus on the latter feature. The former was investigated in the context of supergravity in Ref.~\cite{KYY}, and  it was discussed the implication of the deficient lower multipoles in Refs~\cite{yokoyama,linde, cline, feng}.
Of course both features might be just statistical accidents. However, alternatively, it could be argued that such features give us an insight into the very early universe.  
In the following we show that the hybrid new inflation model is ideal for explaining the low quadrupole, since the density fluctuation generated during the new inflation is naturally
larger than that during the preceding hybrid inflation.

First we review the hybrid inflation model~\cite{hybrid}, which sets the appropriate 
initial condition for the following new inflation. Hereafter we set the gravitational
scale to be unity. The superpotential for the hybrid
inflation is given as
\beq
\label{eq:w_hybrid}
W_H = -\mu^2 S + \lambda S \overline{\Psi} \Psi,
\eeq
which is based on the $U(1)_R$ symmetry, and  the $R$ charges of
$S, \Psi$ and $\overline{\Psi}$ are  $2,1$ and $-1$, respectively.
Assuming the minimal K\"{a}hler potential, the inflationary path is identified
with $\Psi=\overline{\Psi}=0$ and $\sigma > \sigma_c \equiv \sqrt{2}\mu/\sqrt{\lambda}$, where $\sigma$ is the real part of $S$: $\sigma \equiv 
\sqrt{2} {\rm Re}[S]$. 
Since we are concerned with the last a few $e$-folds of
the hybrid inflation, $\sigma$ is relatively small.
In a region of small $\sigma$, radiative corrections cannot be negligible, 
so the potential for $\sigma$ at one-loop order is approximated as~\cite{Dvali}
\beq
\label{eq:pote_for_sigma}
V_H = \mu^4 \left(1+\frac{\lambda^2}{8 \pi^2} \ln \frac{\sigma}{\sigma_c}\right),
\eeq
where we omit the non-renormalizable term since $\sigma$ is relatively small.
The number of
$e$-folds during the hybrid inflation, $N_H$, is related to the inflaton $\sigma$ as
\beq
N_H = \int_{\sigma_c}^{\sigma} d \sigma \frac{V_H}{V_{H}'}
        \simeq \frac{4 \pi^2}{\lambda^2} (\sigma^2-\sigma_c^2),
\eeq
The amplitude of metric perturbations at the horizon crossing is
given by
\beq
\label{eq:amp_hyb}
k^{\frac{3}{2}} \Phi_{k, hybrid} =\left.\frac{V_H^{\frac{3}{2}}}{\sqrt{6}V_{H}'}\right|_{k=aH} =\left.\sqrt{\frac{32}{3}}
\frac{\pi^2 \mu^2}{\lambda^2} \sigma\right|_{k=aH},
\eeq
where $\Phi$ is the gravitational potential, $k$  is a comoving wave number, 
$a$ is the scale factor and $H$ is the Hubble parameter.

Next we consider the new inflation following the hybrid inflation. 
The detailed scenario and application can be found in Refs~\cite{IKY,KSY,KKSY}. 
The superpotential for the new inflation is given by
\beq
W_N=v^2 \phi - \frac{g}{n+1} \phi^{n+1},
\eeq
where $\phi$ has an $R$ charge $2/(n+1)$ and $U(1)_R$ symmetry is dynamically
broken down to $Z_{2n}$ at a scale $v$. Hereafter $v < \mu$ is assumed.
If we identify the real part of $\phi$ as
the inflaton, the potential of $\varphi\equiv \sqrt{2}{\rm Re}[\phi]$ is obtained from the above superpotential  as
\beq
V_N = v^4-\frac{g}{2^{n/2-1}} v^2 \varphi^{n}+\frac{g^2}{2^n} \varphi^{2n},
\eeq
where we assume the minimal K\"{a}hler potential for simplicity.
One of the advantages of this hybrid new inflation is that the initial value of 
$\varphi$ is dynamically set due to the supergravity effect, and it is given by~\cite{KKSY,KYY}
\beq
\varphi_i = \sqrt{\frac{2}{\lambda}} \frac{v^3}{\mu^2}.
\eeq
Thus the $e$-fold number $N_{new}$ is calculated as
\beq
N_{new} = \int_{\varphi_f}^{\varphi_i} d \varphi \frac{V_N}{V_N'}
               \simeq \frac{\lambda^{n/2-1} \mu^{2n-4} v^{-3n+8}}{n(n-2)g},
\eeq
where $\varphi_f$ is the value of $\varphi$ when the slow-roll conditions break, and given by
\beq
\varphi_f = \sqrt{2} \left(\frac{v^2}{n(n-1)g} \right)^{\frac{1}{n-2}}.
\eeq
Likewise, the amplitude of the metric perturbation at the horizon crossing is
\beq
\label{eq:amp_new}
k^{\frac{3}{2}} \Phi_{k, new} =\left.\frac{V_N^{\frac{3}{2}}}{\sqrt{6}V_{N}'}\right|_{k=aH}
 =\left.
\frac{2^{n/2-1} v^4}{\sqrt{6} ng} \varphi^{-n+1}
\right|_{k=aH},
\eeq
and it can be also related to the $e$-folding number $N_{new}$ when evaluated at  comoving
wave number $k_b$ corresponding to the horizon scale at the beginning of the
new inflation:
\beq
k_b^{\frac{3}{2}} \Phi_{k_b, new}=\frac{n-2}{\sqrt{12}} \sqrt{\lambda} \mu^2 v^{-1} N_{new}.
\eeq
Furthermore the spectral index, $n_{new}$, is calculated as
\beq
n_{new}=1-2 \left(\frac{\varphi}{\varphi_f}\right)^{n-2} \simeq 1,
\eeq
where $\varphi \ll \varphi_f$ is assumed in the last equation. Therefore the new inflation adopted here
predicts almost scale invariant power spectrum.

\begin{figure}
    \centering
    \includegraphics[width=8.5cm]{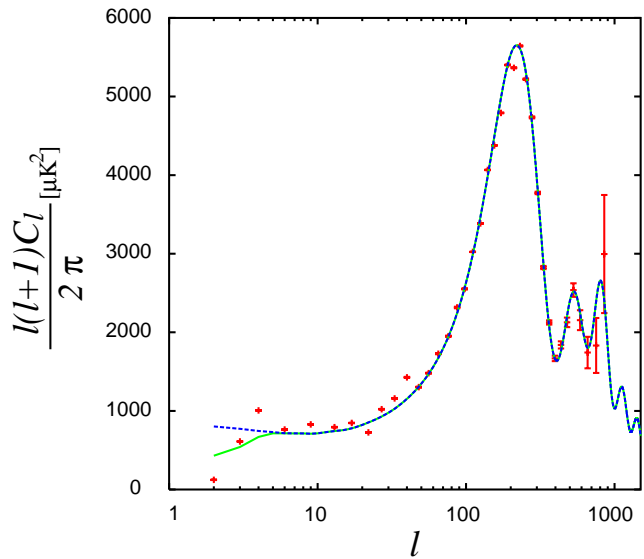}
    \caption{
    The predicted CMB angular power spectrum for our scenario (solid line).
    The WMAP data and the best-fit $\Lambda$CDM model (dotted line)
    are also shown. We have set $k_b = 3 \times 10^{-4}{\rm~ Mpc}^{-1}$,
    $\Omega_bh^2=0.024$, $\Omega_mh^2=0.14$ and  $h=0.72$,  
    where $\Omega_b$ and $\Omega_m$ are density parameters of baryon and 
    non-relativistic matter, respectively and $h$ the Hubble constant in units of 
    100\ km/sec/Mpc, which are suggested from the
    WMAP experiment \cite{WMAP}. 
    }
    \label{fig:cmb}
\end{figure}

We would like to compare the amplitude of the metric perturbation, $k^{\frac{3}{2}} \Phi_{k, new}$, with $k^{\frac{3}{2}} \Phi_{k, hybrid}$ at comoving
wave number $k_b$. Using Eqs.~(\ref{eq:amp_hyb}) and(\ref{eq:amp_new}),
we find that the amplitude of the metric perturbation generated 
during the hybrid inflation is suppressed by a factor
\beq
\kappa \equiv \left. \frac{k^{\frac{3}{2}} \Phi_{k, hybrid}}{k^{\frac{3}{2}} \Phi_{k, new}}
\right|_{k=k_b}=\frac{8 \pi^2 n g \mu^2 \sigma_b \varphi_b^{n-1}}{2^{n/2-1} \lambda^2 v^4},
\eeq
where we defined
\bea
\varphi_b &=& \varphi_i, \non\\
\sigma_b &=& \frac{\lambda}{2 \pi} \sqrt{\frac{2}{3} \ln \frac{\mu}{v}}.
\eea
Using the expression for $N_{new}$, it can be  rewritten as
\beq
\kappa = \frac{8 \pi  \sqrt{ \ln \frac{\mu}{v}}}{\sqrt{3} (n-2)} \frac{v}{\lambda^{3/2} N_{new}}
   \sim \frac{v}{\lambda^{3/2}} \ll 1.
\eeq
Since we require the transitive scale $k_b^{-1}$ be the present horizon size, 
$N_{new}$ must be close to $50$. Note that the hybrid inflation needs not last
for a long time, therefore  $\lambda$ does not have to be much smaller than unity.
The desired amount of the metric perturbation is also obtained if we take
$n=4$, $g=1$, $\lambda = 0.1$, $\mu=3 \times 10^{-6}$ and $v=3.8 \times 10^{-7}$,
for example.  As shown in Refs.~\cite{IY,IKY}, the gravitino mass $m_{3/2}$ and reheating
temperature $T_{RH}$ are related to the model parameters, and are given as
$m_{3/2}\sim170{\rm GeV}$ and $T_{RH} \sim 6.8 {\rm TeV}$ for those exemplified 
values. To demonstrate how the CMB angular power spectrum is deformed
in our scenario, we have performed numerical calculation (see Fig.~\ref{fig:cmb}).
Thus the power spectrum of the density perturbation has the 
desired feature : it is strongly suppressed at the scale larger than $k_b^{-1}$,
while the amount of the density perturbation at smaller scale is reasonable,
reproducing the low quadrupole favored by the WMAP data.
Also note that the power spectrum predicted by the inflation model adopted here is
almost scale invariant, which is consistent with the WMAP results.

In this paper, we have shown that it is possible to suppress the density perturbation
at large scale,  adopting the hybrid new inflation model for a definite discussion.
This mechanism can be responsible for the anomalously low value of the
quadrupole moment of the CMB confirmed by the WMAP. 
Although we have considered one specific inflation model, the strategy is rather
generic. That is to say, in the double inflation model, the density perturbation generated during
the preceding one is somehow suppressed.
For example, it can be also implemented in the curvaton scenario as follows.
We assume such double inflation model that the D-term inflation follows the F-term inflation. 
The curvaton field is assumed to get  the negative Hubble-induced mass term during the
F-term inflation,
while it acts as a massless field during the following D-term inflation. Thus the density perturbation
can have the similar cut-off at large scale, if the D-term inflation lasts for an appropriate
period.

\subsection*{ACKNOWLEDGMENTS}
We acknowledge the use of CMBFAST~\cite{cmbfast} package for
our numerical calculations.
F.T. thanks the Japan Society for the Promotion of Science for
financial support.

\end{document}